\documentclass[usenatbib]{mn2e}

\usepackage{mn2e_journal_style}
\usepackage{graphicx}
\newlength{\figurewidth}
\date{accepted 19 January 2009 - The definitive version is available at
       www.blackwell-synergy.com.}

\title[IGIMF-UV-H$\alpha$ SFRs]{Diverging UV and H$\alpha$ fluxes of star
  forming galaxies predicted by the IGIMF theory}
\author[J.~Pflamm-Altenburg, C.~Weidner and P.~Kroupa]
{Jan~Pflamm-Altenburg$^1$\thanks{email: jpflamm@astro.uni-bonn.de,
    cw60@st-andrews.ac.uk,
    pavel@astro.uni-bonn.de},
  Carsten Weidner$^2$\footnotemark[1]
  and Pavel~Kroupa$^1$\footnotemark[1]\\
  $^1$ Argelander-Institut f\"ur Astronomie, Universit\"at Bonn, 
  Auf dem H\"ugel 71, D-53121 Bonn, Germany \\
  $^2$Scottish Universities Physics Alliance (SUPA), School of Physics and
Astronomy, University of St. Andrews,\\ North Haugh, St. Andrews, Fife KY16
9SS, UK
}

\begin{document}
\maketitle
\begin{abstract}
  Although the stellar initial mass function (IMF) has only been directly
  determined in star clusters it has been manifoldly applied on galaxy-wide
  scales. But taking the clustered nature of 
    star formation into account
    the galaxy-wide IMF is constructed by adding all IMFs of all
    young star clusters leading to an
    integrated galactic initial mass function (IGIMF). 
    The IGIMF is top-light compared to the canonical IMF in star clusters
    and steepens with decreasing total star 
    formation rate (SFR). This discrepancy is marginal for large disk galaxies 
    but becomes significant for SMC-type galaxies and less massive ones.
  We here construct  IGIMF-based relations between
  the total FUV and NUV luminosities of galaxies and the underlying SFR. We
  make the prediction that the H$\alpha$ luminosity of star forming dwarf 
  galaxies decreases faster with decreasing SFR than the UV luminosity.
  This turn-down of the H$\alpha$-UV flux ratio should be evident 
  below total SFRs of 10$^{-2}$~$M_\odot$~yr$^{-1}$.

\end{abstract}
\section{Introduction}
In order to determine the current star formation rate (SFR) of galaxies
certain regions of the electromagnetic spectrum have to be measured which 
have contributions mainly by young stars. For example,
 the H$\alpha$ emission line 
has its origin in the recombination of ionised hydrogen and its intensity 
is a measure for the amount of currently existing short-lived massive
stars. On the other hand, UV luminosities allow a more 
direct access to the young 
stellar population. 

Altough the mass spectrum of UV-luminosity contributing
stars extends to much lower limits than for the H$\alpha$ luminosity
only a massive-star formation rate is measured. In order to estimate
the total galaxy-wide star formation rate an extrapolation to the low-mass
spectrum of newly formed stars has to be done. This extrapolation is 
normally based on the galaxy-wide application of a universal stellar initial
mass function (IMF) leading to linear relations between the total H$\alpha$ and UV
luminosity and the underlying total 
SFR \citep{kennicutt1983a,kennicutt1994a,kennicutt1998b}.   

While the IMF seems to be universal \citep{kroupa2001a,kroupa2002a} its
form has only been determined directly on star cluster scales. The
canonical IMF, $\xi(m)$, is a two-part power-law in the stellar regime,
$dN=\xi(m)\;dm$ being the number of stars in the infinitesimal mass range 
from $m$ to $m+dm$ and $\xi(m)\propto m^{-\alpha_\mathrm{i}}$, $\alpha_1=1.3$ 
for $0.1\le m/M_\odot < 0.5$, $\alpha_2=2.35$ for 
$0.5\le m/M_\odot \le m_\mathrm{max}$, where $m_\mathrm{max}$ is the
maximal stellar mass in a star cluster with the embedded (i.e. birth)
stellar mass $M_\mathrm{ecl}$.

This canonical IMF (or a similar one, e.g. \citealt{chabrier2003a}) 
has traditionally 
been applied on galaxy-wide scales. But most stars form in star clusters
\citep{tutukov1978a,lada2003a}. Thus the
galaxy-wide IMF has to be constructed by adding all young stars of all
young star clusters \citep{weidner2003a,weidner2006a}.
This integrated galactic initial mass function (IGIMF) is steeper than
the canonical IMF in star clusters and steepens with decreasing total SFR
\citep*{weidner2005a,pflamm-altenburg2007d} due to the combination of two
effects:  i) The most  massive star, $m_\mathrm{max}$, in a star cluster is
a function of the
total stellar mass of the young embedded star cluster, $M_\mathrm{ecl}$, \citep{weidner2006a},
and ii) the most massive young embedded star cluster, $M_\mathrm{ecl,max}$, 
is a function of the
total SFR of a galaxy \citep*{weidner2004b}. 
Similar to the IMF in star clusters  the 
embedded cluster mass function (ECMF), which describes the mass spectrum of
newly formed star clusters, follows  a power-law 
distribution function in galaxies  
\citep{lada2003a,weidner2004b,bastian2008a}.  
Due to the $M_\mathrm{ecl,max}$-SFR relation  
the number and total mass ratio of high-mass and low-mass 
young embedded star clusters decreases with decreasing SFR. Because of
the $m_\mathrm{max}$-$M_\mathrm{ecl}$ relation
massive stars are predominantly formed in high-mass clusters, so  
the number and mass fraction of massive stars among all galaxy-wide 
newly formed stars decreases with decreasing SFR, too. Therefore the
IGIMF becomes steeper with decreasing SFR and,
consequently, the fundamental prediction by the IGIMF-theory is that
integrated properties which are sensitive to the presence of massive
stars decrease faster with decreasing SFR than the SFR! We call this basic
phenomenon the IGIMF-effect.

Recently observational evidence has become
available that the galaxy-wide IMF seems to be steeper than  the universal
IMF in star clusters: The Milky Way disk IMF index $\alpha_3\approx 2.7$ (for $m\ge
1\;M_\odot$) is larger than
the canonical Salpeter IMF index of $\alpha_3 \approx 2.35$ 
\citep{kroupa1993a}.  
Using the Scalo index, i.e. $\alpha_3 = 2.7$, \citet{diehl2006a} find a 
consistency between the supernova rate as derived from the Galactic 
$^{26}$Al gamma-ray flux and that deduced from a survey of local O and B
stars, extrapolated to the whole Galaxy \citep{mckee1997a}. 
Unfortunately, a detailed analysis to which order a 
galaxy-wide high-mass IMF slope shallower than Scalo can be ruled out is
not presented in \citet{diehl2006a}.

Additionally, 
\citet{lee2004a} find that low surface brightness galaxies have much steeper
massive-star IMFs than Milky Way-type galaxies. \citet{hoversten2008a}
derived a non-universal stellar IMF in SDSS galaxies based on integrated
photometric properties. Bright galaxies have a high-mass slope of 
$\approx 2.4$, whereas fainter galaxies prefer steeper IMFs. Also,
galaxies with a lower SFR have a steeper IMF than those with a higher SFR
therewith confirming the basic prediction of the IGIMF theory
\citep{weidner2005a}.
Based on
HST-star counts in the SMC-type galaxy 
NGC~4214 \citet{ubeda2007a} derived a massive-star IMF 
slope of  2.8, being steeper than the Salpeter
value. \citet{vanbeveren1983a} had already pointed out that a steepening of
a field IMF compared to the IMF in associations may be expected.

Interestingly, these steeper galaxy-wide IMFs appear to be 
in contradiction
to cosmic IMF studies where the cosmic stellar mass density is compared
with the integrated cosmic star formation rate history. \citet{wilkins2008a}
finds that only a high-mass IMF slope of about 2.15 will put both
cosmic constrains into agreement in the low-redshift
universe. However, their method is an indirect one and depends on 
many other parameters. 

In contrast, the \citet{scalo1986a} high-mass IMF
slope of $\alpha_3=2.7$ implemented by \citet{kroupa1993a} in their adoption
of the standard Galactic-field IMF is based on star counts determining the 
present-day mass function (PDMF) converted into an IMF. 
A steeper than Salpeter
high-mass field IMF slope of the Milky Way is also found by \citet*{reid2002a}
who obtained a slope of 2.5--2.8 at high masses. The conversion of the
PDMF into an IMF requires the assumption of a certain star formation history.
In \citet{reid2002a} a constant star formation rate is assumed based on 
the study by \citet*{gizis2002a}.   
But \citet{elmegreen2006b} demonstrate that true variations in the star formation
rate over times from $\sim{2} \times 10^6$ to $10^9$ yr can produce variations 
in the PDMF slope leading to a wrong conclusion of a varying IMF slope
if the star formation history is assumed incorrectly to vary more smoothly or not at all. 
Thus the assumption of a constant star formation rate in \citet{reid2002a}
may break down on time-scales shorter than $10^{8}$ yr. The Scalo IMF index,
$\alpha_3=2.7$,
is based on an exponentionally changing  star formation rate with a time
constant of 9 to 15~Gyr. A varying time constant between 9 and 15~Gyr does not
influence the derived index significantly. However, short time-scale 
fluctuations as considered by \citet{elmegreen2006b} are not included 
in the analysis by \citet{kroupa1993a}. 
Up to what mass the IMFs derived from the PDMFs in \citet{kroupa1993a}
and \citet{reid2002a} are reliable is an open question, and the local field
IMF inferred from the PDMF therefore needs further 
detailed studies. 

Additionally, galaxy 
evolution models with a steeper than 
Salpeter high-mass IMF slope are in agreement with observed chemical properties
of the solar vicinity \citep{romano2005a} and the low mass-to-light ratios 
of disk galaxies inferred from dynamical arguments \citep{portinari2004a}.
In this view it might be doubtful that the steeper than Salpter
Milky Way field IMFs by \citet{kroupa1993a} and \citet{reid2002a} obtained 
from PDMF star counts are only due to star formation rate variations 
as a galaxy-wide long-term Salpeter high mass IMF index would leed to
expected present day chemical properties of the solar vicinity which 
might be in conflict with local observations.

Chemical evolution models of galaxies \citep*{koeppen2007a}  
have recently shown that
the IGIMF-theory directly leads
to a mass-metallicity relation of galaxies as observed
\citep{tremonti2004a}. Additionally, 
the alpha-element abundances in the IGIMF theory 
vary with the mass of early-type galaxies as is observed \citep{recchi2009a}.
Furthermore, the H$\alpha$ luminosity of a galaxy is related to the presence of
short-lived massive stars such that 
the total H$\alpha$-luminosity of a galaxy scales
non-linearly with the total SFR
\citep{pflamm-altenburg2007d}. 
Consequently, the SFRs as deduced from H$\alpha$ luminosities of
dwarf irregular galaxies are significantly underestimated. Applying the
IGIMF based $L_\mathrm{H\alpha}$-SFR relation to a large sample of star
forming galaxies leads to a linear star formation law for gas rich galaxies,
$SFR\propto M_\mathrm{HI}$,
without any fine-tuning or parameter adjustment, where $M_\mathrm{HI}$ is
the mass of neutral gas. 
It follows that star forming galaxies 
have the same constant gas depletion time scale of
about 2.9~Gyr independently of the galaxy gas mass 
(Pflamm-Altenburg and Kroupa, submitted).

The IGIMF-effect should be more pronounced for integrated properties
which are more sensitive to the most massive stars. Contrary to 
the H$\alpha$ luminosity which depends on the presence of ionizing massive
stars ($\ga 10 M_\odot$), UV luminosities have their main contribution from 
long-lived B stars. Therefore, it is expected that the UV-SFR relation
 is much less SFR- and IGIMF-dependent than the H$\alpha$-SFR relation.

Indeed, observations  of large disk galaxies in the UV-band by GALEX
revealed star formation beyond the
H$\alpha$ cutoff in the very outer disks of galaxies  where the
surface gas density is much lower than the hitherto assumed existing star
formation threshold surface density. This poses a challenge to current
understanding \citep{boissier2007a}. 
This fundamental discrepancy 
is solved naturally in the context of the IGIMF-theory. As the IGIMF-theory
describes the galaxy-wide IMF its local analogon,  
the local integrated galactic initial mass function (LIGIMF), is required  
to address the problem of differing H$\alpha$ and UV star formation rate
surface densities. This has been be
achieved straightforwardly by replacing all galaxy-wide quantities in the
IGIMF-theory by their corresponding surface densities
\citep{pflamm-altenburg2008a}. The LIGIMF-theory perfectly 
reproduces the observed relation between the local H$\alpha$ luminosity 
surface density and the local gas mass surface density and 
the radial H$\alpha$ cut-off in disk galaxies. A basic assumption was that
the UV-luminosity shows no IGIMF-effect and scales linearly with the SFR.
A definite proof for this ansatz was not given but is presented here.

We first describe how the IGIMF based UV-SFR relation is calculated 
(Sec.~\ref{sec_code}), present the relation
(Sec.~\ref{sec_uv-sfr-relation}) and confirm the validity of the ansatz
made in \citet{pflamm-altenburg2008a}. Finally, we calculate 
the expected variation of the H$\alpha$-UV luminosity ratio 
with decreasing total SFR (Sec.\ref{sec_ha_uv}).

\section{Code}
\label{sec_code}
In order to compute the NUV and FUV luminosities of galaxies
for a given SFR we use the second release of the spectral evolution
code P{\sc egase} by \citet{fioc1997a,fioc1999a}. The filter-curves of the
GALEX-FUV and NUV passbands have been obtained from the GALEX webpage. 
The number of points of the tabulated transmission curves has been reduced
to get a simpler description of the filter curves
(Table~\ref{tab_fit_uv_sfr}). 
\begin{table}
\caption{\label{tab_GALEX_filter}GALEX-filter fits.}
\begin{tabular}[t]{cl}
  \hline
  $\lambda$& $T_\mathrm{NUV}(\lambda)$\\
  \hline
     1690&   0.000\\
     1700&   0.025\\
     1750&   0.033\\
     1819&   0.130\\
     1858&   0.191\\
     1904&   0.272\\
     1951&   0.336\\
     2001&   0.450\\
     2050&   0.471\\
     2103&   0.530\\
     2150&   0.595\\
     2200&   0.617\\
     2248&   0.566\\
     2303&   0.527\\
     2350&   0.477\\
     2405&   0.510\\
     2450&   0.535\\
     2556&   0.514\\
     2599&   0.501\\
     2650&   0.457\\
     2700&   0.376\\
     2801&   0.107\\
     2850&   0.035\\
     2900&   0.013\\
     2947&   0.014\\
     3000&   0.019\\
     3010&   0.000\\
\hline
\end{tabular}
\begin{tabular}[t]{cl}
  \hline
  $\lambda$& $T_\mathrm{FUV}(\lambda)$\\
  \hline
     1340 &  0.000 \\ 
     1352 &  0.121 \\
     1370 &  0.177 \\
     1380 &  0.178 \\
     1400 &  0.122 \\
     1427 &  0.257 \\
     1450 &  0.342 \\
     1470 &  0.367 \\
     1480 &  0.368 \\
     1500 &  0.349 \\
     1520 &  0.347 \\
     1530 &  0.329 \\
     1550 &  0.262 \\
     1610 &  0.256 \\
     1651 &  0.159 \\
     1698 &  0.124 \\
     1750 &  0.106 \\
     1799 &  0.012 \\
     1810 &  0.000 \\
     \hline
\end{tabular}

\medskip
The tabulated transmission curves are obtained from the
GALEX-webpage and are piecewisely fitted in order to reduce the
number of segments. The transmission curves $T(\lambda)$
of the NUV and FUV passbands at the wavelength $\lambda$ in 
{\AA} correspond to the number fraction of 
photons transmitted.
\end{table}

The filter-curves are then internally calibrated by P{\sc egase} to the
AB system \citep{oke1974a}. All magnitudes in this paper are absolute
AB magnitiudes unless stated otherwise. An AB-magnitude is related to a
monochromatic flux density, $f_\nu$ in erg~s$^{-1}$~cm$^{-2}$~Hz$^{-1}$,
by 
\begin{equation}
  m_\mathrm{AB}=-2.5\;\log_{10} f_\nu -48.60\,
\end{equation} 
given in \citet{oke1974a}\footnote{Note that there are typos in younger
papers \citep[e.g.]{oke1983a,oke1990a}.}.  An absolute AB-magnitude can then
be converted into a monochromatic luminosity, $L_\nu$ in
erg~s$^{-1}$~Hz$^{-1}$, by
\begin{equation}
  \label{eq_L-M}
  L_\nu = 10^{-0.4(M_\mathrm{AB}-51.60)}\;.
\end{equation}
  
The IGIMFs are calculated for different SFRs 
using equation~11 of \citet{pflamm-altenburg2007d}. The underlying  
IMF in star clusters has the canonical form.
The observed physical upper mass limit 
for stars of about 150~$M_\odot$
\citep{weidner2004a,figer2005a,oey2005a,koen2006a,maiz_apellaniz2006a} has
been reduced to $m_\mathrm{max*}=120 M_\odot$ because the
stellar evolution models included in P{\sc egase} only cover the mass range
from 0.1 to 120~$M_\odot$. This lower physical upper mass limit does not
lead to any falsification of the expected IGIMF-UV luminosities as shown in 
Sec.~\ref{sec_uv-sfr-relation}. 

We choose an 
embedded cluster mass function (ECMF) which is a single-part power law 
with a Salpeter index $\beta_1=2.35$ between 5~$M_\odot$ 
and $M_\mathrm{ecl,max}(SFR)$,  where $M_\mathrm{ecl,max}(SFR)$
is the upper mass limit of the ECMF which is a function of the total SFR
\citep{weidner2004b}. An ECMF with slope 2.35 best reproduces the 
observed relation between the infrared SFRs of galaxies and their most massive
young star cluster \citep{weidner2004b}. Therefore, the IGIMF model with
a Salpter-slope ECMF is the so-called standard model \citep{weidner2005a}. 
This slope is steeper than the ECMF slope of 2 observed for embedded 
low-mass clusters in the solar neighbourhood \citep{lada2003a}. A shallower 
ECMF slope has a higher ratio of high-mass star clusters to low-mass star
clusters and thus leads to a smaller IGIMF effect than an IGIMF model with a
steeper ECMF slope. In order to explore the full range of possible IGIMF
effects we also use a two-part power
law with $\beta_1=1.0$ between 5~$M_\odot$ and 50~$M_\odot$ and $\beta_2=2.0$
between 50~$M_\odot$ and $M_\mathrm{ecl,max}(SFR)$ for the ECMF
to minimise the IGIMF effect.
This is the so-called minimal1-IGIMF 
\citep[table~1 therein]{pflamm-altenburg2007d}. 

The resultant IGIMFs have been  fitted by multi-part power 
laws\footnote{The data files of the fits and an IGIMF-calculation/fit tool 
are available as downloads at \tt{www.astro.uni-bonn.de}} 
and included into P{\sc egase}.

\section{UV-SFR relation}
\label{sec_uv-sfr-relation}
\begin{figure}
\includegraphics[width=\columnwidth]{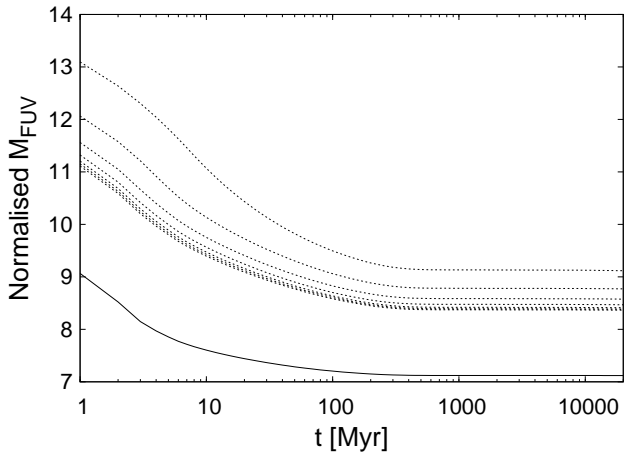}
\caption{\label{fig_fuv_t}Evolution of the normalised FUV-magnitude,
  $M_\mathrm{FUV,norm}$, for a
  constant SFR and a constant metallicity ($Z=0.02$). The solid curve refers to
  the canonical IMF. The dotted curves refer to the standard IGIMF
  corresponding to different total SFRs. From bottom to top: 10$^3$, 10$^2$,
  10, 1, 10$^{-1}$,
  10$^{-2}$, and 10$^{-3}$~$M_\odot$~yr$^{-1}$.}
\end{figure}

\begin{figure}
\includegraphics[width=\columnwidth]{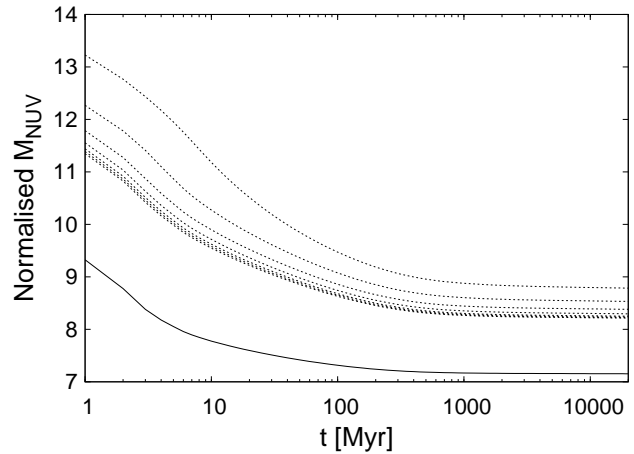}
\caption{\label{fig_nuv_t}Same as Fig.~\ref{fig_fuv_t}, but for NUV.}
\end{figure}

The modified P{\sc egase} code with the new filters and the IGIMF has been 
run with its default values to create a grid of 
galaxy evolution models. Each model has a constant-SFR scenario and no 
evolution of the stellar metallicity. The output by
P{\sc egase} is normalised by the total mass, $M_\mathrm{tot}$, where 
$M_\mathrm{tot}=M_\mathrm{stars}+M_\mathrm{gas}$. The P{\sc egase}
calculations start at $t=0$ with $M_\mathrm{gas}=M_\mathrm{tot}$ and
$M_\mathrm{stars}=0$ and stop at $t=20$~Gyr when 
$M_\mathrm{tot}=\int_{0}^{20\;\mathrm{Gyr}}SFR\;dt=SFR\times 20\;\mathrm{Gyr}$ 
such that $M_\mathrm{tot}$=const
throughout the integrations.
The absolute FUV and NUV magnitudes, $M_\mathrm{UV}$,
and the SFR are obtained from their normalised values by
\begin{equation}
  \label{eq_norm_mag}
  M_{UV} = M_\mathrm{UV,norm} - 2.5\log_{10} M_\mathrm{tot}\;,
\end{equation} 
and
\begin{equation}
  \label{eq_norm_lum}
  SFR = SFR_\mathrm{norm} \; M_\mathrm{tot}\;.
\end{equation}
A relation between the absolute UV magnitude and the SFR can be easily
obtained by combining these two equations and removing the normalisation 
factor/constant. The normalised UV magnitudes evolve with time and 
settle into an 
equilibrium magnitude (Fig.~\ref{fig_fuv_t} and \ref{fig_nuv_t}).
At each time the normalised FUV magnitude is fainter for a lower normalised
SFR as the IGIMF becomes steeper with decreasing SFR, i.e. the fraction of
FUV contributing stars becomes smaller with decreasing SFR.  

\begin{figure}
\includegraphics[width=\columnwidth]{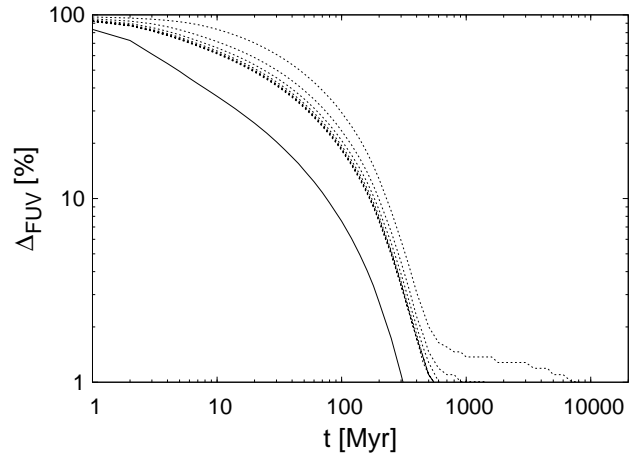}
\caption{\label{fig_fuv_conv_t}Evolution of the relative FUV-luminosity
  difference, $\Delta_\mathrm{FUV}$, compared to the equilibrium luminosity, 
  $L_\mathrm{e}$, at 20~Gyr. The normalised 
  FUV luminosity, $L_\mathrm{t}$, at the
  time $t$ from Fig.~\ref{fig_fuv_t} 
  corresponds to the normalised FUV magnitude, $m_\mathrm{t}$, by 
  $\frac{L_\mathrm{t}}{L_\mathrm{e}}=10^{-0.4
    \left(m_\mathrm{t}-m_\mathrm{e}\right)}$. The plotted value is then calculated by
  equation~\ref{eq_uv_diff}. Note that the value of $\Delta_\mathrm{FUV}$ does
  not depend on the normalisation constant.}
\end{figure}

\begin{figure}
\includegraphics[width=\columnwidth]{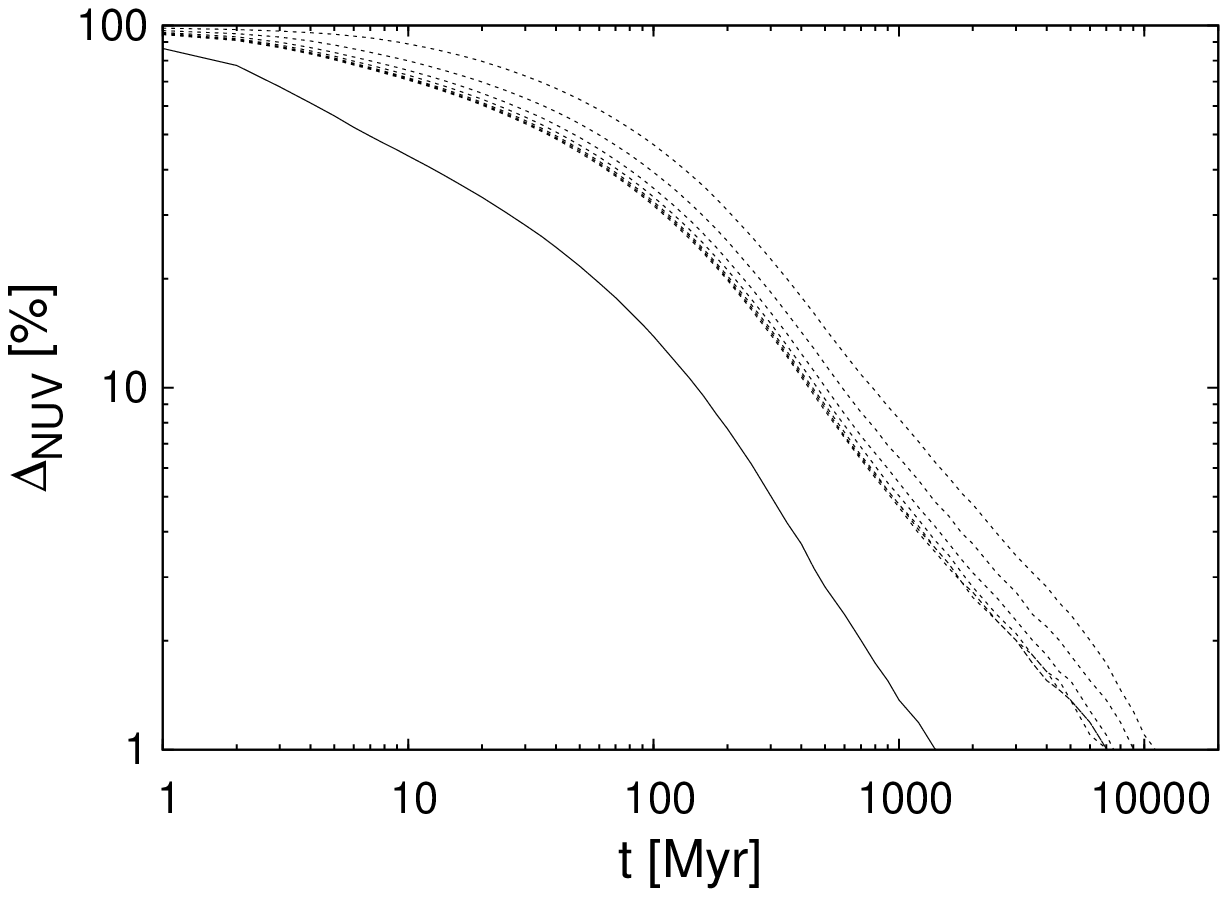}
\caption{\label{fig_nuv_conv_t}Same as Fig.~\ref{fig_fuv_conv_t}, 
but for $\Delta_\mathrm{NUV}$.}
\end{figure}

How fast the UV-magnitude-$t$ curves converge against their equilibrium
value  (evaluated at $t =$~20~Gyr) 
can be seen in Fig.~\ref{fig_fuv_conv_t} and \ref{fig_nuv_conv_t}. 
The normalised UV magnitudes have been converted into a normalised 
luminosity and the relative deviation,
\begin{equation}
  \label{eq_uv_diff}
  \Delta_\mathrm{FUV}=\frac{L(t)-L_\mathrm{e}}{L_\mathrm{e}}
  = \frac{L(t)}{L_\mathrm{e}}-1\;, 
\end{equation} of the
luminosity, $L(t)$, at the time $t$ from the equilibrium luminosity,
$L_\mathrm{e}$, at $t =$~20~Gyr has been plotted as a function of the time.
The value of $\Delta_\mathrm{FUV}$ does not depend on the normalisation 
constant in equation~\ref{eq_norm_mag} as the normalisition constant, 
$M_\mathrm{tot}$, 
in equation~\ref{eq_norm_mag} corresponds to a normalisation factor in 
equation~\ref{eq_norm_lum} which cancels out.
All FUV-IGIMF models differ by less than 2 per cent from the equilibrium value
after 400~Myr, whereas the NUV-IGIMF model are closer than 10 per cent to
the equilibrium value after 1~Gyr. Thus, using the FUV magnitude can only
lead to SFR estimations averaged over $\approx$~400~Myr. Both passbands 
approach the equilibrium value more slowly for lower SFRs. With decreasing SFR
the standard IGIMF becomes steeper and the upper mass limit reduces. 
Thus the mean life-time of UV contributing stars increases and the system 
takes longer to approach the equilibrium state. 

\begin{figure}
  \includegraphics[width=\columnwidth]{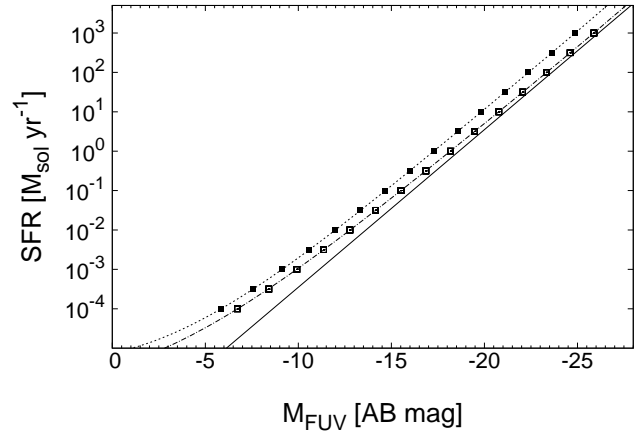}
  \caption{\label{fig_fuv_sfr} The underlying total SFR as a function of the
  equilibrium absolute 
  FUV-magnitude for a constant metallicity ($Z=0.02$) and two
  IGIMF models, standard (black squares) and minimal1 (open squares), 
  and the canonical IMF (solid line). The two dotted curves represent
  polynomial fits (equations~\ref{eq_fit1}--\ref{eq_fit3} and 
  Table~\ref{tab_fit_uv_sfr}) to the IGIMF models.
  The canonical IMF model can be described by equation~\ref{eq_uv_sfr_imf} 
  and Table~\ref{tab_fit_uv_imf}.}
\end{figure}

\begin{figure}
  \includegraphics[width=\columnwidth]{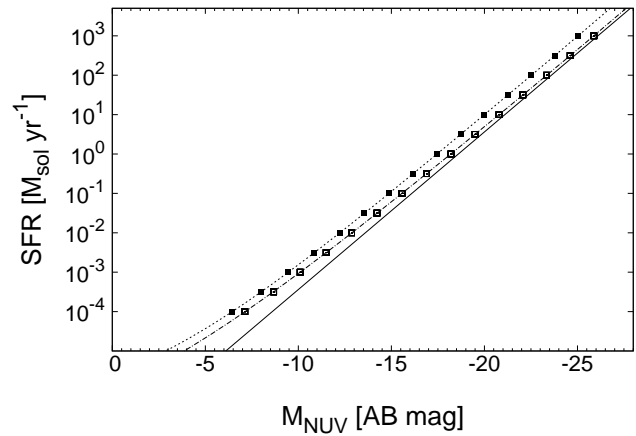}
  \caption{\label{fig_nuv_sfr}Same as Fig.~\ref{fig_fuv_sfr}, but for NUV.}
\end{figure}

The resulting FUV/NUV-SFR relations are shown in Fig.~\ref{fig_fuv_sfr} and
\ref{fig_nuv_sfr} for two IGIMF-models, standard and minimal1, and the
classical linear case, where the IGIMF is identical to the canonical
IMF. Both IGIMF models are fitted by a polynomial of fourth order,
\begin{equation}
  \label{eq_fit1}
  y = a_0+a_1 x + a_2 x^2 + a_3 x^3 +a_4 x^4\;,
\end{equation}
with
\begin{equation}
  \label{eq_fit2}
  x = M_\mathrm{UV}\;[\mathrm{AB\; mag}]
\end{equation}
and 
\begin{equation}
  \label{eq_fit3}
  y = \log_{10}\frac{SFR}{M_\odot\;\mathrm{yr}^{-1}}\;.
\end{equation}
\begin{table*}
\begin{minipage}{126mm}
\caption{\label{tab_fit_uv_sfr}UV-SFR fit (IGIMF).}
{\begin{tabular}[t]{cccccccc}
\hline
  model   & filter &$Z$    & $a_0$ & $a_1$  &  $a_2$ &  $a_3$   &   $a_4$ \\
\hline
std-IGIMF & FUV & 0.0001 & -6.25 & -0.234 & 0.0105 & 0.000311 & 3.57e-06\\
std-IGIMF & FUV & 0.0004 & -6.07 & -0.216 & 0.0116 & 0.000345 & 3.95e-06\\ 
std-IGIMF & FUV & 0.004  & -5.60 & -0.151 & 0.0165 & 0.000509 & 6.08e-06\\
std-IGIMF & FUV & 0.008  & -5.38 & -0.121 & 0.0187 & 0.000585 & 7.08e-06\\
std-IGIMF & FUV & 0.02   & -5.11 & -0.0870 & 0.0214 & 0.000683 & 8.43e-06\\
std-IGIMF & FUV & 0.05   & -4.94 & -0.0705 & 0.0227 & 0.000729 & 9.07e-06\\
min1-IGIMF& FUV & 0.0001 & -6.54 & -0.270 & 0.00734 & 0.000210 & 2.45e-06\\
min1-IGIMF& FUV & 0.0004 & -6.38 & -0.254 & 0.00806 & 0.000226 & 2.59e-06\\
min1-IGIMF& FUV & 0.004  & -5.98 & -0.197 & 0.0120 & 0.000353 & 4.18e-06\\
min1-IGIMF& FUV & 0.008  & -5.74 & -0.162 & 0.0145 & 0.000438 & 5.27e-06\\
min1-IGIMF& FUV & 0.02   & -5.47 & -0.124 & 0.0173 & 0.000533 & 6.52e-06\\
min1-IGIMF& FUV & 0.05   & -5.28 & -0.103 & 0.0187 & 0.000577 & 7.03e-06\\

std-IGIMF & NUV & 0.0001 & -6.90 & -0.325 & 0.00444 & 0.000124 & 1.34e-06\\
std-IGIMF & NUV & 0.0004 & -6.75 & -0.312 & 0.00519 & 0.000144 & 1.55e-06\\
std-IGIMF & NUV & 0.004  & -6.23 & -0.251 & 0.00931 & 0.000273 & 3.11e-06\\
std-IGIMF & NUV & 0.008  & -5.94 & -0.211 & 0.0122  & 0.000368 & 4.33e-06\\
std-IGIMF & NUV & 0.02   & -5.58 & -0.163 & 0.0158  & 0.000494 & 5.98e-06\\
std-IGIMF & NUV & 0.05   & -5.30 & -0.125 & 0.0186  & 0.000590 & 7.22e-06\\
min1-IGIMF& NUV & 0.0001 & -7.13 & -0.362 & 0.00127 & 2.08e-05 & 1.72e-07\\
min1-IGIMF& NUV & 0.0004 & -7.00 & -0.353 & 0.00152 & 2.04e-05 & 9.75e-08\\
min1-IGIMF& NUV & 0.004  & -6.55 & -0.298 & 0.00502 & 0.000128 & 1.39e-06\\
min1-IGIMF& NUV & 0.008  & -6.27 & -0.254 & 0.00817 & 0.000233 & 2.74e-06\\
min1-IGIMF& NUV & 0.02   & -5.96 & -0.209 & 0.0112  & 0.000333 & 4.02e-06\\
min1-IGIMF& NUV & 0.05   & -5.66 & -0.163 & 0.0145  & 0.000438 & 5.28e-06\\
\hline
\end{tabular}}

\medskip
The SFR-absolute-UV-magnitude relations are fitted by a polynomal of 
fourth order according to Eq.~\ref{eq_fit1}--\ref{eq_fit3} for two IGIMF 
models, standard and minimal1, and different metallicities $Z$. 
\end{minipage}
\end{table*}

The resulting co-efficients are listed in Table~\ref{tab_fit_uv_sfr} for 
different metallicities $Z$.
The UV-SFR relation for the unchanging canonical IMF model 
can be expressed by
\begin{equation}
  \label{eq_uv_sfr_imf}
  \frac{SFR}{M_\odot\;\mathrm{yr}^{-1}} = 10^{-0.4 (M_\mathrm{UV}-\mu) - 6}
\end{equation}
and Table~\ref{tab_fit_uv_imf}.
\begin{table}
\caption{\label{tab_fit_uv_imf}UV-SFR fit (IMF).}
\begin{tabular}[t]{ccc}
  \hline
  filter & $Z$ & $\mu$ \\
  \hline
  FUV & 0.0001 &  -4.01 \\
  FUV & 0.0004 &  -3.95 \\
  FUV & 0.004  &  -3.83 \\
  FUV & 0.008  &  -3.75 \\
  FUV & 0.02   &  -3.63 \\
  FUV & 0.05   &  -3.46 \\
  NUV & 0.0004 &  -3.91 \\
  NUV & 0.004  &  -3.75 \\
  NUV & 0.008  &  -3.68 \\
  NUV & 0.02   &  -3.60 \\
  NUV & 0.05   &  -3.49 \\
  \hline
\end{tabular}

\medskip
The SFR-UV-magnitude relations for a galaxy-wide canonical 
IMF for different metallicities $Z$ is expressed by Eq.~\ref{eq_uv_sfr_imf}.
\end{table}

By combining equations~\ref{eq_L-M} and \ref{eq_uv_sfr_imf} a relation 
between the monochromatic luminosity, $L_\nu$ in erg~s$^{-1}$~Hz$^{-1}$,
and the SFR in $M_\odot$~yr$^{-1}$ can be obtained for the case of a 
simple galaxy-wide IMF,
\begin{equation}
  SFR/L_\nu=10^{0.4\mu-26.64}\;.
\end{equation}
For $Z=0.02$ the ratio becomes
\begin{equation}
  SFR/L_\nu=8.09\times 10^{-29}\;,
\end{equation}
which is only a factor 1.3 less than the factor given in \citet{salim2007a}
and a factor 1.7 less than in \citet{kennicutt1998b}. These small 
differences are expected as different stellar population models, different
methods to define a UV-SFR relation, different metallicity
considerations, and slightly different passbands are used.


\begin{figure}
  \includegraphics[width=\columnwidth]{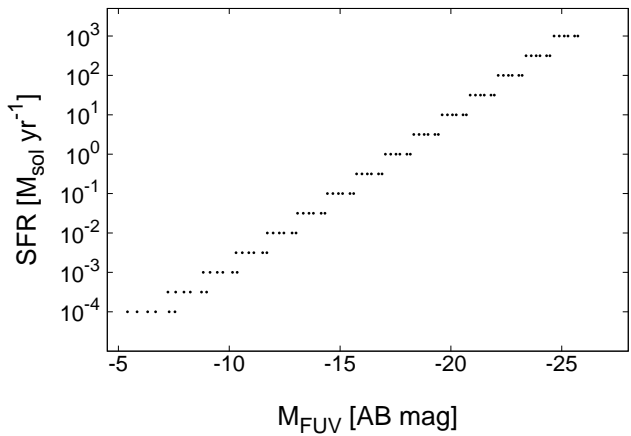}
  \caption{\label{fig_fuv_sfr_z} The SFR as a function of the equilibrium 
    FUV-magnitude for different metallicities (black squares from right 
    to left: $Z$ = 0.0001, 0.0004, 0.004, 0.008, 0.02 and 0.05).}
\end{figure}

Keeping the metallicity constant is justified for calculating the
equilibrium UV magnitude for a given SFR as the equilibrium is 
achieved within $\approx$~400~Myr and the metallicity of a galaxy 
is not expected to change significantly within this period. 
But the absolute constant metallicity may
depend on the given SFR because galaxies with lower SFRs are also 
typically less massive and have lower total metallicities than 
typical galaxies with high SFRs. The standard-IGIMF-SFR is plotted as a 
function of the equilibrium FUV-magnitude for different metallicities 
in Fig.~\ref{fig_fuv_sfr_z}. A very metal poor galaxy ($Z=0.0001$) has
roughly the same FUV-magnitude as a metal rich galaxy ($Z=0.05$) with a
0.5~dex  higher SFR. In particular, applying the FUV-SFR relation for
$Z=0.02$ (Milky Way-type galaxies) on a SMC-type dwarf galaxy with 
$Z\approx 0.004$ would result in a SFR which is
approximately 0.2~dex too high. Such differences are well within any 
observational uncertainties and metallicity differences can be ignored
for the present.

\begin{figure}
  \includegraphics[width=\columnwidth]{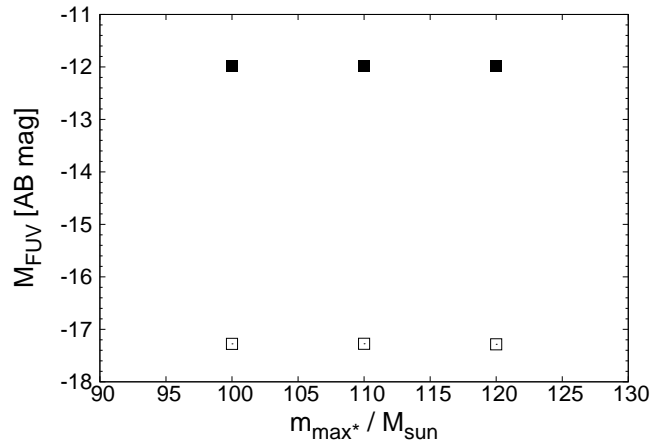}
  \caption{\label{fig_fuv_sfr_mmax} The equilibrium FUV-magnitude as a function
  of the physical upper mass limit, $m_\mathrm{max*}$, for a constant
  metallicity ($Z$ = 0.02) and 
  two different SFRs of 10$^{-2}$ (black squares) and
  1~$M_\odot$~yr$^{-1}$ (open squares) 
  for the case of the standard IGIMF.}
\end{figure}

A second parameter dependence of the UV-SFR relation may be due to
the choice of the physical stellar upper mass limit, $m_\mathrm{max*}$. We here
chose $m_\mathrm{max*}=120$~M$_\odot$ as the stellar evolution models included
in P{\sc egase} do not allow a higher mass limit. But the true upper mass
limit may lie at about 150~$M_\odot$
\citep{weidner2004a,figer2005a,oey2005a,koen2006a,maiz_apellaniz2006a}. To
test for a possible physical 
upper-mass-limit dependence we calculate the FUV equilibrium magnitude for 
the standard IGIMF and a constant metallicity of $Z=0.02$ but with 
$m_\mathrm{max*}$=100, 110 and 120~$M_\odot$. As can be seen in 
Fig.~\ref{fig_fuv_sfr_mmax} there is no
influence on the FUV magnitude when varying the physical stellar upper 
mass limit above 100~$M_\odot$. Because the NUV magnitude is less sensitive
to the presence of high-mass stars than FUV the non-dependence on the
physical upper mass limit holds for NUV, too.

\section{Comparing the UV-- and H$\alpha$--SFR}
\label{sec_ha_uv}
As already mentioned the 
UV-SFR relation is expected to be very much less sensitive to the IGIMF effect
and thus much closer to  linearity than the corresponding 
H$\alpha$-SFR relation.
In order to demonstrate this we predict the result a classical observer
would get when comparing UV and H$\alpha$ based SFRs obtained with a linear 
luminosity-SFR relation.
 
For a given true total SFR the H$\alpha$ luminosity
and the total FUV magnitude are calculated for the
IGIMF theory. The FUV magnitude is calculated using the FUV-SFR
relation constructed in Section~\ref{sec_uv-sfr-relation}. Although 
the P{\sc egase} output provides H$\alpha$ luminosities,
the H$\alpha$-SFR relation from \citet{pflamm-altenburg2007d} is used
instead. P{\sc egase} does not include stellar evolution models for stars more
massive than 120~$M_\odot$. But the physical stellar upper mass limit
is thought to lie around 150~$M_\odot$ 
\citep{weidner2004a,figer2005a,oey2005a,koen2006a,maiz_apellaniz2006a}
which is taken into account in the H$\alpha$-SFR relation in 
\citet{pflamm-altenburg2007d}. 
As shown in the previous section the exact value of the physical upper mass
limit does not have a noticeable influence on the FUV-SFR
relation. Contrary, the H$\alpha$-SFR relation does appreciably  depend on
the physical stellar upper mass limit as discussed below. 
The H$\alpha$ luminosity is
then reconverted into a SFR using the classical linear 
H$\alpha$-SFR relation \citep{kennicutt1994a,kennicutt1998b}. 
The IGIMF FUV-magnitude is reconverted into a SFR using the linear 
relation described by  equation~\ref{eq_uv_sfr_imf} 
based on the invariant canonical IMF. 

Both, the theoretical H$\alpha$- and the theoretical FUV-luminosity 
refer to produced radiation, i.e. extinction is not taken into account. 
The reason is that both produced luminosities and therefore
their ratio is only a function of the total
SFR. The apparent luminosities and luminosity ratios indeed depend 
on different extinction effects, thus do not depend on the total SFR but e.g.
on the dust content of the target galaxy, the inclination of the target
galaxy, and the Galactic latitude and longitude. Thus two galaxies with 
identical SFRs and produced H$\alpha$- and FUV-luminosities can have
different observed luminosities. The correction for
Galactic and internal extinction has to be done individually for
each observed target galaxy.

\begin{figure}
  \includegraphics[width=\columnwidth]{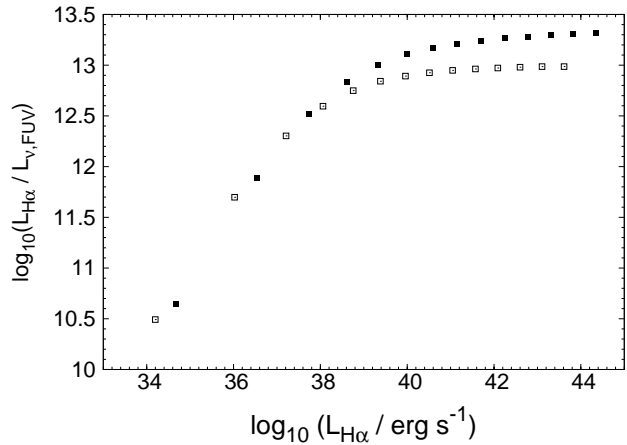}
  \caption{\label{fig_halpha_fuv_lum} 
    The expected H$\alpha$-/UV-luminosity ratio as a function of
    the H$\alpha$ luminosity for two different underlying IGIMF models, 
    standard (open squares) 
    and minimal1 (filled squares).
  }
\end{figure}

Fig.~\ref{fig_halpha_fuv_lum} shows the expected ratio of the total
H$\alpha$ and FUV luminosity as a function of the total H$\alpha$
luminosity for two different IGIMF models, standard (open squares)
and minimal1 (filled squares), and a constant metallicity ($Z=0.02$).   

\begin{figure}
  \includegraphics[width=\columnwidth]{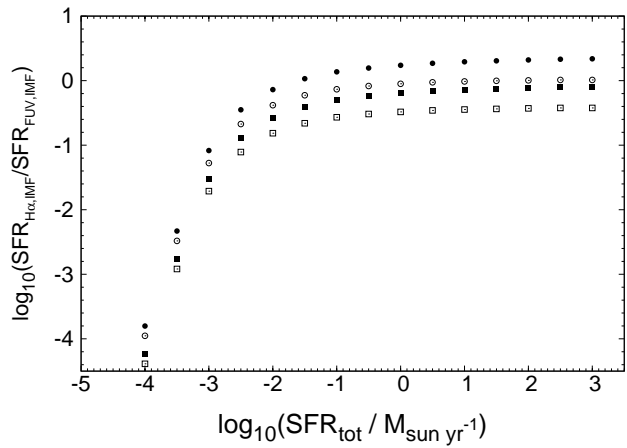}
  \caption{\label{fig_halpha_fuv} The expected H$\alpha$-SFR/UV-SFR ratio when
  the luminosities are converted into SFRs using classical linear relations
  for two different underlying IGIMF models, standard (open symbols) 
  and minimal1 (filled symbols), and two different
  linear H$\alpha$-SFR relations, the canonical-IMF based relation from 
  \citet{pflamm-altenburg2007d} (squares) and the widely used 
  Salpeter-IMF based relation from \citet{kennicutt1994a} (circles).
}
\end{figure}

Fig.~\ref{fig_halpha_fuv} shows the expected
UV-H$\alpha$ SFR-ratio as a function of the true underlying SFR for a
constant metallicity ($Z=0.02$). The ratio is plotted for two IGIMF models,
standard (open symbols) and minimal1 (filled symbols), and two different
linear H$\alpha$-SFR relations, the invariant canonical-IMF based relation from 
\citet{pflamm-altenburg2007d} (squares) and the widely used 
Salpeter-IMF based relation from \citet{kennicutt1994a} (circles). 

Below a total SFR of about 10$^{-2}$~$M_\odot$~yr$^{-1}$ the SFR 
based on H$\alpha$ flux and an invariant IMF
should start to dramatically decrease faster 
than the SFR based on UV flux and an invariant IMF. 

IGIMF model details such 
as different slopes of the ECMF
have only a minor effect on the position and the strength of the H$\alpha$-UV
turn-down. Furthermore, at first sight it is expected that the classical
H$\alpha$-SFR should be always lower than the classical UV-SFR:
Consider a galaxy which populates its young stellar content 
according to the IGIMF scenario. The number of O and B stars in this galaxy
is smaller than expected if the galaxy-wide IMF were identical to the canonical 
IMF for the same SFR. In order to get the same total UV and H$\alpha$ 
flux for an assumed galaxy-wide invariant IMF the calculated SFR has to be
smaller than the true SFR. As the H$\alpha$ flux decreases faster than the 
UV flux with decreasing SFR the SFR based on H$\alpha$ flux and an invariant
IMF is always smaller than the SFR based on UV flux and an invariant IMF.

But as shown in the previous section, varying the physical
stellar upper mass limit has no influence on the FUV magnitude. On the 
other hand, the situation is 
different for the H$\alpha$ luminosity as it depends
significantly on the number of very massive stars. 
The IGIMF model includes an IMF which
can be populated up to a physical stellar upper mass limit of 150~$M_\odot$
in agreement with recent 
observations
\citep{weidner2004a,figer2005a,oey2005a,koen2006a,maiz_apellaniz2006a}. 
When using the linear H$\alpha$-SFR relation based on the canonical IMF
with a physical stellar upper mass limit of 150~$M_\odot$ from
\citet{pflamm-altenburg2007d} the H$\alpha$-SFR/UV-SFR ratio converges as
expected against unity with increasing SFR. But when calculating the H$\alpha$
based SFR with the widely used Salpeter-IMF based relation by 
\citet{kennicutt1994a}, 
which has an upper mass limit of 100~$M_\odot$, the classical H$\alpha$
based SFRs can be higher than the UV-based SFRs, because the H$\alpha$
contribution of stars more massive than 100~$M_\odot$ has to be balanced
by an artificial higher SFR when taking only stars less massive than 
100~$M_\odot$ into account.
  
\section{Conclusions}
Based on the IGIMF theory, which takes the clustered nature of star 
formation into account, we have constructed a relation between 
the UV-luminosity of a galaxy and the underlying SFR. In agreement
with previous assumptions and expectations we confirm that the UV
luminosities of galaxies are much less sensitive to the IGIMF effect than
the H$\alpha$ luminosity.  
Furthermore, the definite prediction by the IGIMF theory is that 
around a total SFR of about 10$^{-2}$~$M_\odot$~yr$^{-1}$ the classical 
IMF based H$\alpha$ SFR should start to decrease faster 
than the classical IMF based UV SFR.
The position and the strength of the turn-down is nearly
IGIMF-model independent. This prediction can be tested
by UV-observations of dwarf irregular galaxies with total SFRs
in the range of 10$^{-4}$--10$^{-2}$~$M_\odot$.

\vspace{1cm}
We would like to thank Samuel Boissier for helpful comments on this
paper.
J.~P-A. acknowledegs partial support through DFG grant KR1635/20. C.W.
acknowledegs support through European Commission Marie Curie Research
Training Grant CONSTELLATION (MRTN-CT-2006-035890)
\bibliographystyle{mn2e}
\bibliography{galaxy-evolution,star-formation,imf,cmf,OB-star,star-cluster,stellar-spectra}

\end{document}